\documentclass[aps,pra,twocolumn,superscriptaddress]{revtex4}
\usepackage{graphicx}
\usepackage{dcolumn}

\begin{document}

\title{Explicit construction of nonadiabatic passages for stimulated Raman
transitions}

\author{Hong Cao}
\affiliation{Center of Theoretical Physics, College of Physics, Sichuan University, Chengdu 610065, China}

\affiliation{School of Material Science and Engineering, Chongqing Jiaotong
University, Chongqing 400074, China}

\author{Shao-Wu Yao}
\affiliation{Center of Theoretical Physics, College of Physics, Sichuan University, Chengdu 610065, China}

\author{Li-Xiang Cen}
\email{lixiangcen@scu.edu.cn}
\affiliation{Center of Theoretical Physics, College of Physics, Sichuan University, Chengdu 610065, China}

\begin{abstract}
We propose a scheme which can produce desired
nonadiabatic passages for the stimulated Raman transition in three-level systems.
The state transfer in the protocol is realized
by following the evolution of the dynamical basis
itself and no additional coupling field is required.
We also investigate the interplay between the present nonadiabatic protocol and
the shortcut to adiabaticity.
By incorporating the latter technology, we show that alternative
passages with less occupancy of the intermediate level could be designed.
\end{abstract}

\maketitle

\section{Introduction}

Stimulated Raman adiabatic passage (STIRAP) is an efficient technique for
robust coherent population transfer in atomic and molecular systems, which
has been extensively investigated over the past decades
\cite{raman0,raman1,raman2,raman3n,raman3,raman4}. It can induce transitions between two levels
that has the same parity, for which the direct coupling via electric dipole
radiation is forbidden. Specifically, the STIRAP applies two laser pulses,
the pump and Stokes, to induce the coupling between each of the two levels
and a common intermediate level under the condition of the two-photon
resonance. The desired population transfer is realized through the adiabatic
evolution of the dark state, which conventionally assumes a superposition
form of the initial and the final target states.

Based on the transitionless tracking algorithm \cite{counter,counter2,berry}, the shortcut to
adiabaticity \cite{shortcut1,shortcut2,shortcut3,shortcut33,shortcut4,shortcut5,shortcut6}
has been exploited to speed up the evolution of the STIRAP. In the initial proposals
\cite{shortcut2,shortcut3} of the stimulated Raman shortcut-to-adiabatic
passage (STIRSAP), the application of a compensating microwave field which
couples the initial and target states is required to counteract the
detrimental nonadiabatic effect. This additional microwave field indicates a
critical disadvantage, especially concerning that the direct coupling might
be unfeasible in practical atomic levels with forbidden transition. In the
subsequent proposals \cite{shortcut4,shortcut5}, it is displayed that one can mimic
the desired population transfer of the STIRSAP protocol through modifying
the pump and Stokes pulses, which removes the coupling term associated with
the microwave field so as to avoid the former drawback. Strictly, in this
modified STIRSAP scheme the evolution of the wavefunction (so called as the
``dressed state" in Ref. \cite{shortcut4}) differs from that in the former
one by a rotating transformation $V(t)$. The validity of the scheme then
relies on the boundary condition of $V(t)$, that is, $V(t)$ should be a null
operation at the initial (ending) time instant.

Successful design of the laser pulses for the STIRSAP is critically
constrained by the above boundary condition of the dressed-state
transformation. For example, this condition is not satisfied when the pump
and the Stokes fields assume the commonly used Gaussian pulses \cite
{shortcut4,shortcut5}. Note that the rotating angle of $V(t)$ is correlated
to the aforementioned microwave field that should have been applied in the
initial STIRSAP protocol. The specified boundary condition of $V(t)$ can
be fulfilled only when the strength of this additional interaction,
or equally speaking, the nonadiabatic effect induced by the initial pump and
Stokes fields, should be negligible on the boundary.
This actually requires that the driving protocol should satisfy the adiabatic condition
at that time instant.
At this stage, a promising design may rest on (but not limited to) the
prerequisite that the system should possess discrete energy levels at
the initial (ending) instant of the driving pulses.

On the other hand, valuable results have been obtained recently in
understanding the nonadiabatic dynamics generated by some particular types
of quantum driven models \cite{model1,model2,model3,model4}.
It is shown that the nonadiabatic effect in some of the cases can play a
positive role for the population transfer. For example, in the
tangent-pulse-driven model with the matching frequency and amplitude, the
nonadiabatic effect not only will not lead to unwanted transitions but also
can suppress the error caused by the truncation of the field pulse \cite
{model1}. This feature of the nonadiabatic driving has also been found in a
modified Landau-Zener model \cite{model2} and a special Allen-Eberly model
\cite{model3}. Motivated by these results, it is natural to ask whether
there exists such kind of nonadiabatic passages that can be exploited
directly to realize the stimulated Raman process.

In this paper we report the finding of a specific driving scheme via which
the nonadiabatic passages of the stimulated Raman transition can be explicitly
constructed. The scheme applies to the $\Lambda $-type three-level system with
the one-photon resonance in which the Stokes laser pulse can be of arbitrary
analytical form but the pump pulse should be matching
with the Stokes one. Several driving protocols generated by the scheme are illustrated
and there various features for the population transfer are characterized.
Moreover, we explore the interplay of the present scheme with the STIRSAP
and show how to reconstruct the nonadiabatic passages within the framework of
the shortcut to adiabaticity. Incorporation with the latter technology enables us to
design alternative nonadiabatic passages with less occupancy of the intermediate level.

\section{Stimulated Raman nonadiabatic passages with one-photon resonance}

\subsection{Description of the driving scheme}

Consider a three-level system with the $\Lambda $ configuration shown in
Fig. 1. The states $|1\rangle $ and $|3\rangle $ are ground or metastable
levels, which are coupled to the excited level $|2\rangle $ via the laser
pulses, the pump pulse and the Stokes pulse, respectively. The Hamiltonian
of the system under the rotating wave approximation can be written as 
$H_{\rm {tot}}(t)=H_{\rm {free}}+H_{\rm {int}}(t)$ with the free Hamiltonian 
$H_{\rm {free}}=\sum_{i=1}^3E_i\sigma _{ii}$ and the interaction term
\begin{eqnarray}
H_{\rm{int}}(t) &=&\Omega _p(t)(\sigma _{12}e^{i\omega _pt}+\sigma
_{21}e^{-i\omega _pt})  \nonumber \\
&&+\Omega _s(t)(\sigma _{23}e^{-i\omega _st}+\sigma _{32}e^{i\omega _st}),
\label{hamil}
\end{eqnarray}
in which $\sigma _{ij}=|i\rangle \langle j|$ $(i,j=1,2,3)$ and $\Omega _{p,s}(t)$ describe
the Rabi frequencies of the pump and Stokes pulses, respectively. Under the
condition of the one-photon resonance $\omega _p=E_2-E_1$ and $\omega
_s=E_2-E_3$, one obtains the Hamiltonian in the interaction picture
\begin{equation}
H(t)=\Omega _p(t)(\sigma _{12}+\sigma _{21})+\Omega _s(t)(\sigma
_{23}+\sigma _{32}).  \label{hamil1}
\end{equation}
To implement fast population transfer from the state $|1\rangle $ to 
$|3\rangle $, we propose a nonadiabatic protocol in which the laser pulses
satisfy
\begin{equation}
\Omega _p(t)=\frac 12\Omega _s(t)\sec [\frac 12\int_{t_0}^t\Omega _s(\tau
)d\tau ]  \label{passage}
\end{equation}
with the envelope of the Stokes laser $\Omega _s(t)$ being an arbitrary
analytical function over $t\in (t_0,t_f)$. As is shown in the below, when
the integral of the intercepted pulse $\int_{t_0}^t\Omega _s(\tau )d\tau
\equiv 2\vartheta (t)$ goes from $0$ to $\pi $, complete population transfer
$|1\rangle \rightarrow |3\rangle $ can be realized by the protocol in a
nonadiabatic manner.

\begin{figure}[t]
  \includegraphics[width=7cm]{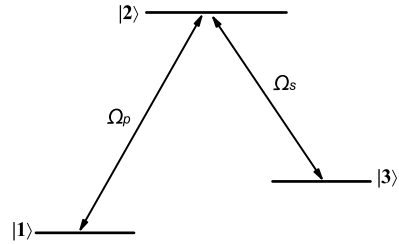}
  \caption{Schematic of the level structure of the three-level system, where $\Omega_p$ and $\Omega_s$
  denote the Rabi frequencies of the pump and Stokes laser pulses, respectively.}
  \label{fig1}
\end{figure}

To resolve the dynamics of the above stimulated Raman process, we note
that the described model possesses a dynamical invariant \cite{lewis1,lewis2}
\begin{eqnarray}
I(t) &=&\sin ^2\vartheta (t)(\sigma _{12}+\sigma _{21})+\cos \vartheta
(t)(\sigma _{23}+\sigma _{32})  \nonumber \\
&&-i\sin \vartheta (t)\cos \vartheta (t)(\sigma _{13}-\sigma _{31}),
\label{inva}
\end{eqnarray}
which satisfies
\begin{equation}
\partial _tI(t)=-i[H(t),I(t)].  \label{LRequ}
\end{equation}
It is recognized that the three operators $K_x\equiv \sigma _{12}+\sigma
_{21}$, $K_y\equiv i(\sigma _{13}-\sigma _{31})$ and $K_z\equiv \sigma
_{23}+\sigma _{32}$ satisfy the commutation relation $[K_\alpha ,K_\beta
]=i\epsilon _{\alpha \beta \gamma }K_\gamma $. By recording $I(t)=\vec{\chi
}(t)\cdot \vec{K}$, Eq. (\ref{LRequ}) is readily verified through the
following equations of the components
\begin{eqnarray}
\dot{\chi}_1(t) &=&-\Omega _s(t)\chi _2(t),  \label{comp1} \\
\dot{\chi}_2(t) &=&\Omega _s(t)\chi _1(t)-\Omega _p(t)\chi _3(t),
\label{comp2} \\
\dot{\chi}_3(t) &=&\Omega _p(t)\chi _2(t).  \label{comp3}
\end{eqnarray}
The eigenvalues of $I(t)$ are given by $\lambda _0=0$ and $\lambda _{\pm
}=\pm 1$, and the eigenstate $|e_0(t)\rangle $ associated with the zero
eigenvalue $\lambda _0$ is obtained as
\begin{equation}
|e_0(t)\rangle =\cos \vartheta |1\rangle -i\sin \vartheta \cos \vartheta
|2\rangle -\sin ^2\vartheta |3\rangle .  \label{bright}
\end{equation}
It is seen that the initial state $|1\rangle $ correlates exclusively with
the basis state $|e_0(t)\rangle $, so the wavefunction will evolve along
this ``dressed state'' for the time being. Since the corresponding
eigenvalue $\lambda _0=0$, the Lewis-Riesenfeld phase \cite{lewis1,lewis2}
accumulated during the evolution is zero for $|e_0(t)\rangle$.
As $\vartheta (t_f)\rightarrow \frac \pi 2$, the state transfer
$|1\rangle \rightarrow |3\rangle $ is achieved up to a minus sign.

A particular feature one can recognize from the above stimulated Raman protocol is that
at the initial $t=t_0$ there is $\Omega_p(t_0)=\frac 12\Omega_s(t_0)$. This is distinctly
different from the delayed pulse sequence of the adiabatic passage, which
indicates that the nonadiabatic effect
plays a decisive role in the present protocol. It also reveals that
the Hamiltonian and the dynamical invariant are not commutative
at $t=t_0$: $[H(t_0),I(t_0)]\neq 0$, which does not accord with the condition assumed
in Ref. \cite{shortcut2}. Secondly, as $t\rightarrow t_f$ the asymptotic population
on the target state $|3\rangle $ is specified by $P(t)=\sin ^4\vartheta
(t)$. It indicates that the protocol is less sensitive to the cutoff error of the field pulses
than the adiabatic protocol. In detail, suppose
that the field pulses are truncated at $t=t_{fc}$ with the Rabi frequencies $\Omega_p(t_{fc})$
and $\Omega_s(t_{fc})$. Denote by $\delta \equiv \arctan\frac{\Omega _s(t_{fc})}{\Omega _p(t_{fc})}$
the deviation of the pulses. The population on $|3\rangle$ at $t=t_{fc}$ is then given by
\begin{equation}
P(t_{fc})=\sin^4\vartheta (t_{fc})=(1-\frac 14\tan ^2\delta)^2.
\end{equation}
As $\delta$ is much less that $1$, it is not difficult to verify that there is always
$P(t_{fc})\geq \cos^2\delta$. That is to say, comparing with the conventional STIRAP,
the nonadiabatic effect in the present protocol will reduce the loss of fidelity caused
by the truncation.

\subsection{Typical nonadiabatic passages}

The above scheme offers an explicit way to construct stimulated Raman nonadiabatic
passages via which the population transfer $|1\rangle\rightarrow|3\rangle$ can be realized.
We present some typical examples in the below.

Example 1. The Rabi frequency of the Stokes laser is set to be a
constant. According to Eq. (\ref{passage}), there are
\begin{equation}
\bigg\{
\begin{array}{l}
\Omega _p(t)=\frac \nu 2\sec \frac{\nu t}2, \\
\Omega _s(t)=\nu,
\end{array}
\label{example1}
\end{equation}
in which the time $t$ goes from $t=0$ to $t\rightarrow\frac \pi \nu$.
The dynamical invariant and the zero-eigenvalue dress state are specified by
\begin{eqnarray}
I(t) &=&\sin ^2\frac{\nu t}2(\sigma _{12}+\sigma _{21})+\cos \frac{\nu t}
2(\sigma _{23}+\sigma _{32})  \nonumber \\
&&-\frac i2\sin \nu t(\sigma _{13}-\sigma _{31})  \label{invar1}
\end{eqnarray}
and
\begin{equation}
|e_0(t)\rangle =\cos \frac{\nu t}2|1\rangle -\frac i2\sin (\nu t)|2\rangle
-\sin ^2\frac{\nu t}2|3\rangle ,  \label{dark1}
\end{equation}
respectively. In this proposal the pump laser assumes a chirped pulse and
the duration of the pulse $\nu t_f\approx \pi$ is much shorter than that
of the usual adiabatic protocol.

Example 2. The passage is described by
\begin{equation}
\bigg\{
\begin{array}{l}
\Omega _p(t)=\nu, \\
\Omega _s(t)=2\nu \mathrm{sech}(\nu t),
\end{array}
\label{example2}
\end{equation}
in which $\Omega _p(t)$ of the pump laser is a constant. The two Rabi frequencies above
are verified to satisfy Eq. (\ref{passage}) in view that the half of the
integration of $\Omega_s(t)$ gives rise to $\vartheta (t)=\arctan [\sinh
(\nu t)]$. The eigenstate $|e_0(t)\rangle$ of the dynamical invariant
is obtained as
\begin{equation}
|e_0(t)\rangle =\mathrm{sech}(\nu t)|1\rangle -i\tanh (\nu t)\mathrm{sech}%
(\nu t)|2\rangle -\tanh ^2(\nu t)|3\rangle .  \label{dark2}
\end{equation}
The field pulses in this protocol are defined in an infinite time domain
and the truncation is inevitable. We define an effective pulse duration
$t\in(t_0, t_{fc})$ in which
$t_{fc}$ is defined such that the population
on $|3\rangle$ reaches $P(t_{fc})\approx 0.9999$. For the current example
there is $\nu t_{fc}\approx 1.8\pi$.

In principle, unlimited amount of the nonadiabatic passages could be constructed by the scheme.
Besides the above two, some other examples are displayed in Table I, including the field pulses,
evolution of the population, and their effective pulse duration.

\begin{table*}[t]
\caption{Various nonadiabatic passages and their pulse durations generated by the scheme.}
\label{twod}%
\begin{ruledtabular}
\begin{tabular}{c|llll|ll}
~~    &  Stokes Pulse & Pump Pulse &   Target State Pop. & $\nu t_{fc}$ &  $\Omega_{s0}(t)$ & $\Omega_{p0}(t)$ \\ \hline
$(a)$  & $  \Omega_s=\nu	$	&	$\Omega_p=\frac {\nu} {2}\sec \frac {\nu t}{2}$	 & $\sin^4\frac {\nu t}{2}$	&	$\approx \pi$	& $\frac{\nu}{2}$ & $\frac{\nu}{2}\tan(\frac{\nu t}{2})$ \\
$(b)$  & $  \Omega_s=2\nu \mathrm{sech}(\nu t)$ &	$\Omega_p=\nu$	 & $\tanh^4(\nu t)$	&	$\approx 1.80\pi$	& $\nu \mathrm{sech}(\nu t)$ & $\nu\tanh(\nu t)$  \\
$(c)$  & $  \Omega_s=\frac {2\nu}{\sqrt{1-\nu^2 t^2}}	$	&	$\Omega_p=\frac {\nu} {1-\nu^2 t^2}$	& $\nu^4 t^4$ &	$\approx 1$	&  $\frac {\nu}{\sqrt{1-\nu^2 t^2}}$ & $\frac {\nu^2 t}{1-\nu^2 t^2}$ \\
$(d)$  & $  \Omega_s=\nu^{2}t	$	&	$\Omega_p=\frac {\nu^{2}t} {2}\sec \frac {\nu^{2} t^{2}}{4}$	 & $\sin^4\frac {\nu^{2} t^{2}}{4}$	&	$\approx 0.80\pi $	& $\frac{\nu^{2}t}{2}$ & $\frac{\nu^{2}t}{2}\tan(\frac{\nu^{2} t^{2}}{4})$\\
$(e)$  & $  \Omega_s=4\nu^{2} t \mathrm{sech}(\nu^{2} t^{2})$ &	$\Omega_p=2\nu^{2}t$	 & $\tanh^4(\nu^{2} t^{2})$	&	$\approx 0.76\pi$	& $2\nu^{2}t \mathrm{sech}(\nu^{2} t^{2})$ & $2\nu^{2}t\tanh(\nu^{2} t^{2})$ \\
$(f)$  & $  \Omega_s=2\nu e^{\nu t}	$	&	$\Omega_p=\nu e^{\nu t}\sec(e^{\nu t}-1)$	& $\sin^{4}(e^{\nu t}-1)$ &	$\approx 0.30\pi$	&  $\nu e^{\nu t}$ &$\nu e^{\nu t}\tan(e^{\nu t}-1)$
\end{tabular}
\end{ruledtabular}
\end{table*}

\section{Interplay with the shortcut to adiabaticity}

\subsection{Stimulated Raman shortcut-to-adiabatic passage and the dressed-state
transformation}

To be specific, let us review the STIRSAP protocol with the one-photon
resonance of which the initial Hamiltonian in the interaction picture reads
\begin{equation}
H_0(t)=\Omega _{p0}(t)(\sigma _{12}+\sigma _{21})+\Omega _{s0}(t)(\sigma
_{23}+\sigma _{32}).  \label{hamil0}
\end{equation}
In general, the sequence of the delayed pulse interactions of the pump and the Stokes
lasers are implemented in the counterintuitive order so that the Rabi frequencies
satisfy $\Omega
_{s0}(t_0)\gg \Omega _{p0}(t_0)$ and $\Omega
_{s0}(t_f)\ll \Omega _{p0}(t_f)$ at the initial and the ending time $t_{0,f}$, respectively.
For the adiabatic evolution the state transfer $|1\rangle\rightarrow|3\rangle$ can be realized
along the dark state $|d(t)\rangle =\cos \theta_0(t)|1\rangle -\sin \theta _0(t)|3\rangle $
with $\theta _0(t)=\arctan \frac{\Omega _{p0}(t)}{\Omega _{s0}(t)}$.
Following the shortcut-to-adiabatic technology \cite{shortcut2,shortcut3,shortcut33,shortcut4,shortcut5},
the nonadiabatic effect of the evolution could
be cancelled by introducing a compensating
microwave field $H_{cd}(t)=i\dot{\theta}_0(t)(\sigma _{13}-\sigma _{31})$
and the corrected Hamiltonian
\begin{equation}
H_{\mathrm{corr}}(t)=H_0(t)+H_{cd}(t)  \label{hamilc}
\end{equation}
can drive the system along the eigenstate $|d(t)\rangle $ of the initial
$H_0(t)$ in a nonadiabatic manner. Note that
$H_{cd}(t)$ represents a direct coupling between the levels $|1\rangle $ and
$|3\rangle $ which might be unavailable for the control setup. To overcome this
drawback, one can replace the above $H_{\mathrm{corr}}(t)$ by an alternative
one \cite{shortcut4,shortcut5}
\begin{equation}
\tilde{H}_{\mathrm{corr}}(t)=V(t)H_{\mathrm{corr}}(t)V^{\dagger
}(t)-iV(t)\partial _tV^{\dagger }(t),  \label{hamilcc}
\end{equation}
in which $V(t)=e^{i\phi (t)(\sigma _{23}+\sigma _{32})}$ accounts for a
rotating transformation. When the rotating angle
is set as $\phi (t)=\arctan \frac{\dot{\theta}
_0(t)}{\Omega _{p0}(t)}$, the interacting term with respect to the direct coupling between
$|1\rangle $ and $|3\rangle $ will disappear in the corrected Hamiltonian $\tilde{H}_{\mathrm{corr}}(t)$.
The corresponding dynamical basis of $\tilde{H}_{\mathrm{corr}}(t)$ relates to $|d(t)\rangle $ via: $|\tilde{d}
(t)\rangle =V(t)|d(t)\rangle $. Therefore, as long as the boundary conditions
$|\tilde{d}(t_{0,f})\rangle =|d(t_{0,f})\rangle$ are
satisfied, the desired population transfer $|1\rangle\rightarrow|3\rangle$ could be
realized via the evolution of the dressed state $|\tilde {d}(t)\rangle$.

It is worthy to mention that the condition $|\tilde{d}(t_0)\rangle =|d(t_0)\rangle$
is always fulfilled for the STIRSAP since the equality $V(t_0)|1\rangle=|1\rangle$
holds for any given $\phi(t_0)$. That is to say, the boundary of $V(t_0)$ is irrelevant
but only the condition $V(t_f)=I$ is required in the design of the protocol.
As will be shown in the below, this is the case (i.e., $V(t_0)\neq I$)
when we reconstruct the nonadiabatic passages proposed in the previous section
within the framework of the STIRSAP.

\subsection{Reconstructing the nonadiabatic passages via the shortcut-to-adiabatic
technology}

Let us move to consider the issue of constructing the nonadiabatic passages described in
Sec. II through the STIRSAP scheme. The goal now is to determine
reversely the initial Hamiltonian $H_0(t)$ based on the known target
Hamiltonian $\tilde{H}_{\mathrm{corr}}(t)=H(t)$ specified in Eqs. (\ref
{hamil1}) and (\ref{passage}). Note that the dynamical invariant of the system
$\tilde{H}_{\mathrm{corr}}(t)$ is known to be the $I(t)$ of Eqs. (\ref{inva})
and the corresponding dressed state $|\tilde{d}(t)\rangle \equiv|e_0(t)\rangle $
is shown in Eq. (\ref{bright}).
It is readily seen that a rotating transformation $V^\dagger(t)=e^{-i\phi (t)(\sigma
_{23}+\sigma _{32})}$ with $\phi (t)=\frac \pi 2-\vartheta (t)$ can transform
$|\tilde{d}(t)\rangle $ to the dark state $|d(t)\rangle $:
\begin{equation}
V^\dagger (t)|\tilde{d}(t)\rangle =\cos
\vartheta (t)|1\rangle -\sin \vartheta (t)|3\rangle \equiv|d(t)\rangle.
\label{targetd}
\end{equation}
Then the inverse transformation on Eq. (\ref{hamilcc}) gives rise to
\begin{eqnarray}
H_{\mathrm{corr}}(t) &=&V^{\dagger }(t)\tilde{H}_{\mathrm{corr}%
}(t)V(t)-iV^{\dagger }(t)\partial _tV(t)  \nonumber \\
&=&\Omega _p(t)\sin \vartheta (t)K_x+\frac{\Omega _s(t)}2(K_y+K_z).
\label{hamilc2}
\end{eqnarray}
By comparing the above expression with Eq. (\ref{hamilc}),
one recognizes that
\begin{equation}
H_0(t)=\Omega _p(t)\sin \vartheta (t)(\sigma _{12}+\sigma _{21})+\frac{%
\Omega _s(t)}2(\sigma _{23}+\sigma _{32})  \label{hamil0s}
\end{equation}
and
\begin{equation}
H_{cd}(t)=\frac 12\Omega _s(t)K_z\equiv i\dot{\vartheta}(t)(\sigma
_{13}-\sigma _{31}).  \label{hamilcd}
\end{equation}
More explicitly, the form of the laser pulses of $H_0(t)$ can be expressed as
\begin{equation}
\bigg\{
\begin{array}{l}
\Omega _{p0}(t)=\frac {\Omega_s(t)} {2}\tan \vartheta (t),  \\
\Omega _{s0}(t)=\frac {\Omega_s(t)} {2},
\end{array}
\label{initial0}
\end{equation}
where $\vartheta_0(t)\equiv\theta_0(t)=\arctan\frac {\Omega_{p0}(t)}{\Omega_{s0}(t)}$.

So far, we have completed the reconstruction of the nonadiabatic driving scheme
through the STIRSAP, that is, the state transfer
process realized by the control Hamiltonian of Eqs. (\ref{hamil1}) and (\ref{passage})
can be understood as
the transitionless algorithm tracking the evolution of
the instantaneous eigenstate $|d(t)\rangle$ of the Hamiltonian $H_0(t)$,
corrected by a dressed-state transformation $V(t)=e^{i\phi (t)(\sigma_{23}+\sigma _{32})}$.
As $\phi (t)$ here goes from $\phi (t_0)=\frac \pi 2$ to $\phi (t_f)=0$, it confirms
the aforementioned statement that only the boundary condition $V(t_f)=I$ is necessary.
Furthermore, for the concrete nonadiabatic passages specified by Eqs. (\ref{example1}) and (\ref{example2}),
one obtains
\begin{equation}
\bigg\{
\begin{array}{l}
\Omega _{p0}(t)=\frac \nu 2\tan \frac{\nu t}2, \\
\Omega _{s0}(t)=\frac \nu 2,
\end{array}
\label{example10}
\end{equation}
and
\begin{equation}
\bigg\{
\begin{array}{l}
\Omega _{p0}(t)=\nu \tanh (\nu t), \\
\Omega _{s0}(t)=\nu ~\mathrm{sech}(\nu t),
\end{array}
\label{example20}
\end{equation}
respectively. For more examples, the corresponding expressions
of $\Omega _{p0}(t)$ and $\Omega _{s0}(t)$ are
displayed in the last column of Table I.

\section{Strategy to reduce the intermediate-level occupancy}

Differing from the original STIRAP in which the population transfer is realized
along the dark state, the occupancy on the excited level $|2\rangle$ will occur
in the present nonadiabatic protocol when the system evolves along the dressed state
$|\tilde{d}_0(t)\rangle$. The Raman passages proposed here therefore will be sensitive
to the decay of the intermediate level, which is somewhat similar to the
bright-STIRAP \cite{bSTIRAP1,bSTIRAP2,bSTIRAP3}. Note that the detrimental effect
on the bright-STIRAP due to the dissipation of this auxiliary intermediate level
has ever been estimated in Refs. \cite{dissp1,dissp2,dissp3,dissp4,dissp5,dissp6}.
For comparison, it is expected that the detrimental effect should be less serious
in the present scheme since the pulse durations of the passages here (see Table I)
are much shorter than those of the adiabatic passages.

Besides, the intermediate-level occupancy of the present
scheme can be reduced by following the approach of Ref. \cite{shortcut4}, that is,
one can adjust the dressed state $|\tilde{d}(t)\rangle$ by constructing alternative control
Hamiltonians. Essentially, this strategy makes use
of the multiplicity of the control Hamiltonian of the tracking algorithm when aiming at the
desired state evolution \cite{berry}. That is, one can use the series of Hamiltonians
$\{H_0^{\prime }(t)=\eta (t)H_0(t)\}$ to replace $H_0(t)$ specified by
Eqs. (\ref{hamil0s}) and (\ref{initial0}) with $\eta(t)$ an adjustable factor.
Subsequently, the shortcut-to-adibatic protocol should give rise to
\begin{equation}
H_{\mathrm{corr}}^{\prime }(t)=\eta (t)H_0(t)+H_{cd}(t),  \label{hamilco}
\end{equation}
with $H_{cd}(t)$ being the same as in Eq. (\ref{hamilcd}).
Following the protocol, one can find that the rotating angle of the dressed-state
transformation $V(t)$ should change as
\begin{equation}
\phi (t)\rightarrow \phi ^{\prime }(t)=\arctan [\eta ^{-1}(t)\cot \vartheta
(t)].  \label{angle2}
\end{equation}
Accordingly, the new Hamiltonian $\tilde{H}_{\mathrm{corr}}^\prime (t)$ and the dressed
state can be formulated as
\begin{equation}
\tilde{H}_{\mathrm{corr}}^\prime (t)=\frac {1}{2}\Omega_s\sqrt{1+\eta^2\tan^2\vartheta}K_x+(\frac{\eta}{2}\Omega_s-\dot{\phi}^\prime)K_z,
\label{dressH}
\end{equation}
and
\begin{eqnarray}
|\tilde{d}^{\prime }(t)\rangle &=&e^{i\phi ^{\prime }(t)(\sigma
_{23}+\sigma _{32})}|d(t)\rangle
\label{dress2} \\
&=&\cos \vartheta |1\rangle -i\sin \vartheta
\sin \phi ^{\prime }|2\rangle -\sin \vartheta \cos \phi ^{\prime }|3\rangle , \nonumber
\end{eqnarray}
respectively.

\begin{figure}[t]
  \includegraphics[width=7.5 cm]{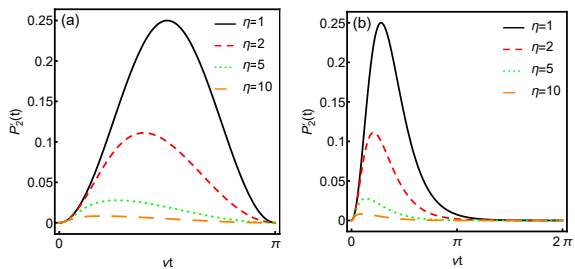}
  \caption{Reduction of the intermediate-level occupancy by modifying the dressed state,
  see. Eqs. (\ref{dress2})-(\ref{ratio}). (a) The first example (a) shown in Table I.
  (b) The second example (b) shown in the Table.}
  \label{fig2}
\end{figure}

Now, it is straightforward to see that the population on the intermediate level $|2\rangle $
changes from $P_2(t)=\sin^2\vartheta (t)\cos^2\vartheta (t)$ [see Eq. (\ref{bright})]
to
\begin{equation}
P_2^{\prime}(t)=\frac{\cos^2\vartheta (t)}{\eta^2(t)+\cot^2\vartheta(t)}.
\label{pop22}
\end{equation}
As long as $|\eta(t)|>1$, the above strategy will lead to reduction of the population on $|2\rangle$
with the ratio
\begin{equation}
\frac {P_2^{\prime }(t)}{P_2(t)}=\frac 1{\eta ^2(t)\sin ^2\vartheta (t)+\cos ^2\vartheta (t)}.
\label{ratio}
\end{equation}
For the case that $\eta$ is a constant, the maximal inhibition rate is achieved at a time point
with $\vartheta =\frac \pi 4$, wherein the maximal occupancy on the level $|2\rangle$ just right happens.
The value of the rate there is obtained as $P_2^{\prime }/P_2=2/(1+\eta ^2)$.
As is illustrated in Fig. 2, sizable reduction of the occupancy of the intermediate level
could be achieved for the driving protocol by adjusting the factor $\eta$.

\section{Conclusion}

In summary, we have proposed a nonadiabatic driving scheme to realize the stimulated Raman process.
The scheme applies no auxiliary coupling field but only the pump and the Stokes lasers to the $\Lambda$-type
system with one-photon resonance.
The nonadiabatic effect is shown to play a decisive role in the state transfer
protocol and some typical Raman nonadiabatic passages generated by the scheme
are illustrated. We further investigate the interplay between the present scheme and the STIRSAP
based on the transitionless tracking algorithm. By incorporating the latter technology, we show that
alternative nonadiabatic passages with less occupancy of the intermediate level could be constructed.


\end{document}